\documentclass[a4paper]{spie}  

\usepackage{amsmath,amsfonts,amssymb}
\usepackage{graphicx}
\usepackage{physics}
\usepackage{dcolumn}
\usepackage{bm,bbm}
\usepackage{subcaption}
\usepackage{capt-of}
              
\newcommand{\eref}[1]{Eq.~(\ref{#1})}
\newcommand{\sref}[1]{Sec.~\ref{#1}}
\newcommand{\fref}[1]{Fig.~\ref{#1}}

\title{A practical compact source of heralded single photons for simple detection LIDAR}

\author{Mateusz P. Mrozowski}
\author{John Jeffers}
\author{Jonathan D. Pritchard}
\affil{Department of Physics, University of Strathclyde, John Anderson Building, \\107 Rottenrow East, Glasgow G4 0NG, United Kingdom}
\authorinfo{M.M.: E-mail: mateusz.mrozowski@strath.ac.uk}

\begin{document} 
\maketitle
\begin{abstract}
Optical quantum technologies such as quantum sensing, quantum cryptography and quantum computation all utilize properties of non-classical light, such as precise photon-number and entangled photon-pair states, to surpass technologies based on the classical light. 
A common route for obtaining heralded single photons is spontaneous four-wave mixing in optical fibers, allowing for a well-defined spatial mode, for high efficiency integration into optical fiber networks. 
These fibers are typically pumped using large, commercial, pulsed lasers requiring high-power ($\sim$10 W) pump lasers and are limited to $\sim$MHz repetition rate.
Here we propose a cost-efficient, compact and mobile alternative. Photon pairs at 660 nm and 960 nm will be created using four-wave mixing in commercial birefringent optical fiber, pumped using transform limited picosecond pulses with GHz repetition rates derived from a 785 nm CW laser diode using cavity-enhanced optical frequency comb generation. 
The pulses are predicted to have average power of 275 mW, a peak power of $>$40 W, and predicted photon yield of $>$2000 pairs detected per second.
This design will be later utilized to implement a quantum illumination scheme based on a coincidence count between idler and signal photons – instead of joint measurement between signal and idler. 
This will allow for quantum advantage over classic LIDAR without the requirement for maintaining an interferometric stability in free space.
  
\end{abstract}


\section{INTRODUCTION}
\label{sec:intro}  
Photon pairs are an essential resource, either as factorisable states of pure photons for applications such as quantum computation with linear optics \cite{Kok2007}, and quantum simulations using photons \cite{PhysRevLett.118.130503}, or as entangled pairs used to enhance timing measurements \cite{Harris2007}, improve clock synchronization \cite{Giovannetti2001}, cancel dispersion in interferometers \cite{Franson2009} and for quantum illumination \cite{England2019,Lloyd1463,PhysRevLett.101.253601,Shapiro2009,PhysRevA.80.052310}.

Whilst entangled light sources enable quantum illumination to achieve a theoretical detection enhancement over classical illumination\cite{Lloyd1463}, developing a detection scheme able to utilise this regime in practice - without sensitivity to decoherence or need for an interferometric stability on a length scale of an optical wavelength, was so far possible only within an optical fiber \cite{Zhang2015}. Recent proposals have shown that quantum illumination using heralded rather than entangled photons can also be exploited in LIDAR schemes\cite{Frick2020,Yang2021}, in which simple detection based on 
coincidence measurements using single photon detector modules has been demonstrated to enable rejection of thermal background and robustness against spoofing\cite{Lloyd1463}.

Two common routes to obtain heralded photons are Spontaneous Parametric Down Conversion (SPDC) in non-linear crystals \cite{Kaneda:16}, and Spontaneous Four-Wave Mixing (SFWM) in birefringent optical fibers \cite{Smith:09}. 
The latter offers advantages in the form of a well defined spatial mode for high efficiency coupling into optical fiber networks developed for quantum information protocols \cite{PhysRevA.79.030303,matthews2009manipulating,smith2009phasecontrolled}, as well as the lower material cost of optical fibers, compared to the cost of non-linear photonic crystals. 
Recent studies have shown that using SFWM in commercial grade polarisation-maintaining single-mode optical fibers provides a high purity and wavelength tuneable source of heralded single photons, adaptable to both visible\cite{Smith:09} and telecoms wavelengths\cite{Lugani2020}.

In typical experiments, photon pairs are obtained by pumping birefringent fibers using bulky, high power ($\sim10$ W) pulsed laser systems, which come with significant price-tag ($\sim\pounds100$k), and a fixed repetition rate ($\sim1-100$ MHz). 
Here we propose an alternative low cost and compact, heralded photon source, offering a high repetition rate ($\sim$ GHz), and a tuneable pulse width ($1-10$ ps) that will allow for temporal control of the generated photons. 
This source uses an optical frequency comb generator (OFCG) based on an intra-cavity electro-optical modulator (EOM) to convert a CW pump into a mode-locked pulse train \cite{Kovacich:00,250392,doi:10.1063/1.1654403}. A dual-cavity design is adopted to achieve a high efficiency output \cite{Macfarlane:96}, overcoming the major limitation of a single cavity OFCG sources.
We estimate a photon pair detection rate of $>2000$ photon pairs/s following simulation of our pump source, currently limited by the damage threshold of the intra-cavity EOM.





\section{Optical Frequency Comb Generation}
\label{OFCG}
\subsection{Single cavity OFCG}
\label{SCEOM}
\begin{figure}[b]
    \centering
    \includegraphics[width=\textwidth]{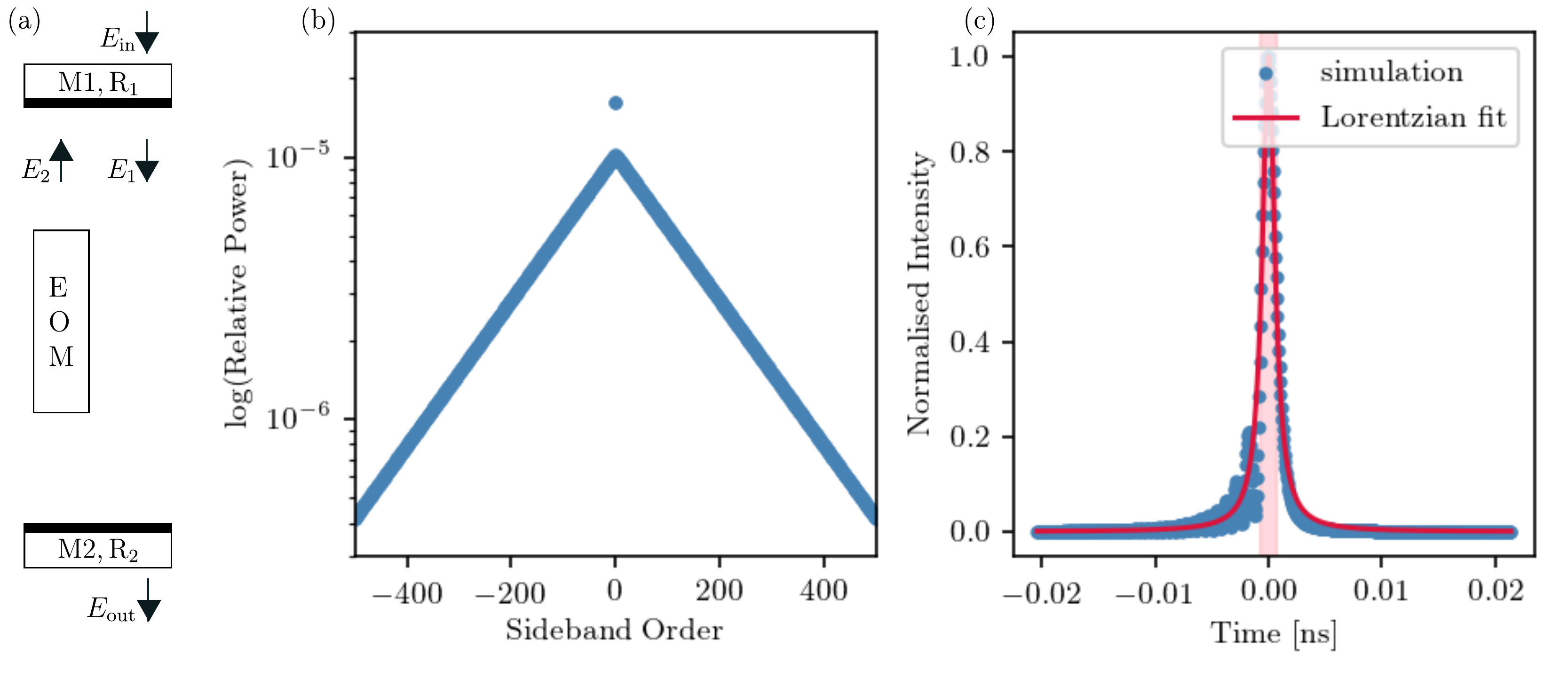}
    \vspace{0.01\linewidth}
    \caption{(a) Model of single cavity OFCG. Here we are using an approximation, in which light passes through the EOM only once per round trip, while the crystal is assumed to be twice as long and modulation depth is doubled. (b) Spectral output of the single cavity OFCG. Efficiency of the system $\eta = 0.3\%$. (c) Temporal output of the single cavity OFCG with a FWHM pulse $\Delta\tau = 1.3$ ps as indicated by the red shaded area. The asymmetry is an artifact of the crystal dispersion. }
    \label{1}
\end{figure}

A simple OFCG source involves use of an EOM inside an optical cavity driven resonantly with the cavity free spectral range (FSR) to generate a comb of phase-coherent sidebands that convert a CW pump laser into a train of short optical pulses with a repetition rate equal to twice the modulation frequency.
The output electric field can be calculated by modelling the round-trip propagation as illustrated in \fref{1}(a) 
using the mathematical model presented in \sref{sec:matha}, resulting in an output field $\Vec{E}_{\mathrm{out}}$ equal to
\begin{equation}
     \Vec{E}_{\mathrm{out}} = t_{\mathrm{c}}\hat{\phi}\left(\frac{\phi_k}{2}\right)\times M(\delta)\times t_2 \left(\left[ \mathbbm{1} - r_1r_2t_{\mathrm{c}}^2\hat{\phi}(\phi_k)\times M(2\delta) \right]^{-1} \times t_1 \Vec{E}_{\mathrm{in}}\right),
     \label{eq:eq22}
\end{equation}
\noindent where $\hat{\phi}(\phi_k)$ and $M(2\delta)$ are the full round trip phase change operator and full round trip coupling matrix respectively, $\delta$ is the modulation depth in radians, $\phi_k$ is the round-trip phase change of $k^\mathrm{th}$ sideband, $t_i$ and $r_i$ are transmittance and reflectivity of the mirror $M_i$, 
$t_{\mathrm{c}}$ is transmission through the EOM crystal, 
and $\Vec{E}_{\mathrm{in}}$ is the single frequency optical field represented as a vector with central mode ($k = 0$) having an intensity equal to 1.
This provides the electric field represented in the frequency domain from which the temporal response can be obtained using a Fourier transform. The frequency spectrum of the output exhibits an exponential decay in the electric field as a function of sideband order\cite{250392}, therefore the pulse shape will be Lorentzian of the form \cite{doi:10.1063/1.1654403}
\begin{equation}
    \label{erling}
    E(t) \propto \left[\frac{1}{1+(4f_m\delta Ft)^2}\right]E_{\mathrm{in}},
\end{equation}
\noindent where $f_m$ is the modulation frequency, and $F$ is the cavity finesse. The pulse width $\Delta \tau_p$ at FWHM intensity can then be approximated as\cite{doi:10.1063/1.1654403}
\begin{equation}
    \label{erling2}
    \Delta\tau_p \approx \frac{\sqrt{\sqrt{2}-1}}{2f_m\delta F}.
\end{equation}

In the system consisting of an EOM placed in the cavity, the dispersion comes from both the dispersive effect of the electro-optic crystal, as well as the dispersion of the cavity mirrors. The latter can be neglected because it is small compared to the dispersive effect of the crystal. The full round trip phase change per sideband is given by \cite{477736}
\begin{equation}
    \phi_k = \frac{k^2}{2}\mathrm{GVD}\left(2\pi f_m\right)^2L_c, 
    \label{eq:eq17}
\end{equation}
\noindent where $k$ represents the sideband order, $f_m$ is the frequency with which the EOM is modulated, $\mathrm{GVD}$ is the group velocity dispersion of the material, and $L_c$ is the crystal's length. 
The phase difference between the carrier and the sidebands generated by the EOM increases quadratically with sideband order. 
Eventually that difference in phase will be large enough for the sideband to no longer be resonant with the optical cavity, because of the destructive interference that light mode will be suppressed, and in effect the frequency comb will be truncated. This dispersion limit produces a sharp cut-off in the optical frequency comb spectrum, limiting it's span to \cite{477736}

\begin{equation}
    \label{eq:1.3.1}
    \Delta f = \frac{\left[2\left(-\beta + \delta \right)G\right]^{1/2}}{\pi},
\end{equation}
\noindent where $\beta$ is the normalized detuning between the input laser frequency $f_o$ and the nearest cavity resonance $f_r$, defined as $\beta = \left[\pi\left(f_o - f_r\right)\right]/\mathrm{FSR}$, $\delta$ is modulation depth, and $G$ is related to the EOM material dispersion by $G = 2\pi c/D\lambda_o^2L_c$, where $L_c$ is the crystal length, $\lambda_o$ is the input laser wavelength in vacuum and $D=\left.\frac{\lambda_o}{c}\frac{\partial^2n}{\partial\lambda^2}\right|_{\lambda = \lambda_o}$ is the material dispersion which can be related to the group velocity dispersion by use of \cite{Kovacich:00} $\mathrm{GVD} \cong D\lambda_o^2/2\pi c$.

Using this model we extract typical performance for a single cavity system using the following parameters.
Mirrors 1 and 2 were set to equal reflectivity $R_1=R_2=0.99$, with the EOM modulation depth $\delta=\pi/2$, and $f_m= 2.39$ GHz. 
Due to the light passing twice through the EOM, it experiences a modulation of $2\delta$ meaning for $\delta > \pi/2$ the light becomes resonant with the next cavity mode leading to an unstable output rate and uneven intervals between output pulses\cite{Kovacich:00,Xiao:08}.

The EOM is modelled as an $L_c=20$ mm magnesium-doped lithium tantalate crystal (MG:LiTaO$_3$) chosen for its linear group velocity dispersion ($\mathrm{GVD}=308$ fs$^2$/mm\cite{Moutzouris:11}) at the central wavelength of $\lambda_0= 785$ nm and assumed to have an anti-reflective (AR) coating with $R=0.01\%$ on each facet.
The output electric field generated in the single cavity can be seen in \fref{1}(b). As expected, in the frequency domain we can observe an exponential relationship between sidebands leading to a Lorentzian temporal output pulse, as shown in \fref{1}(c). The resulting FWHM pulse width $\Delta\tau_p=1.3$ ps and the efficiency of the system $\eta = 0.3\%$ at a repetition rate of $2f_m = 4.78$ GHz.


\subsection{A Coupled Cavity OFCG}
\label{CCOFCG}
\begin{figure}[t!]
    \centering
    \includegraphics[width=\textwidth]{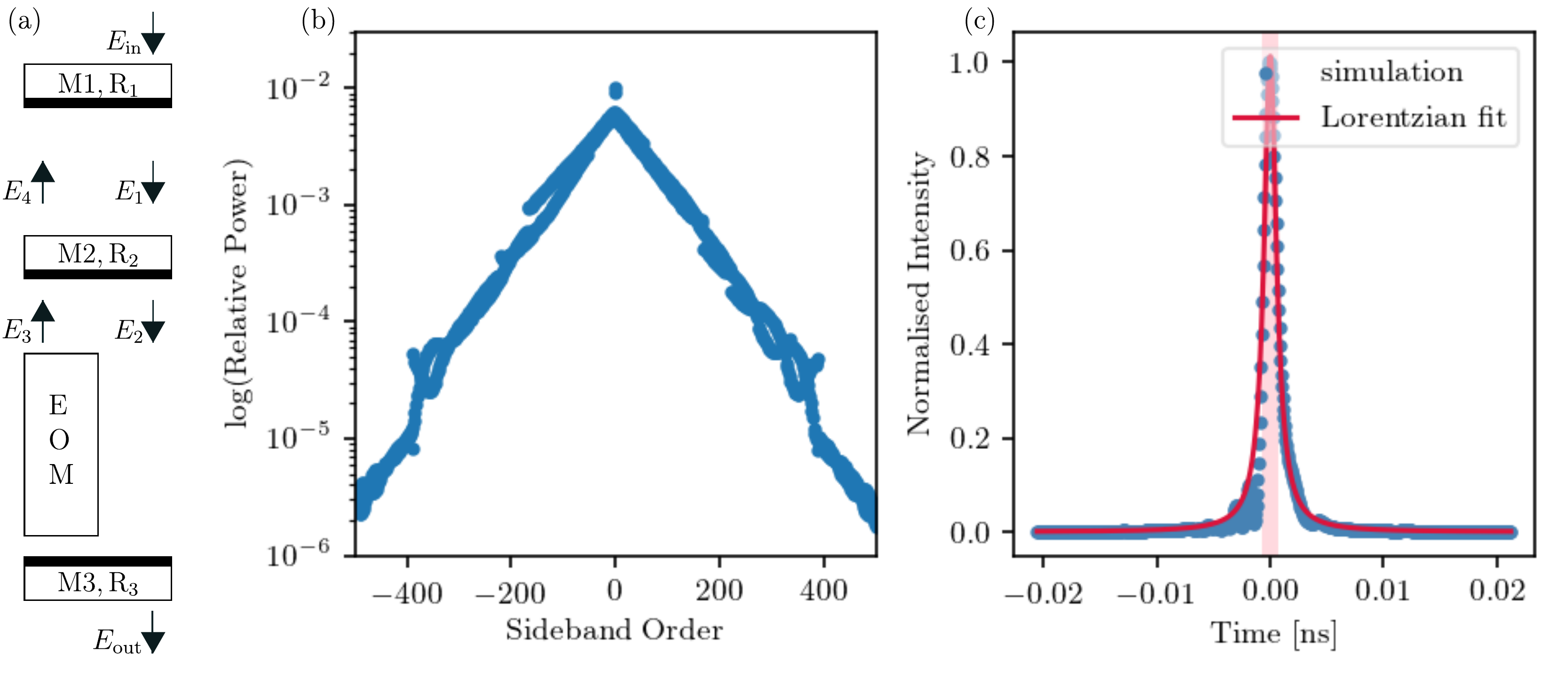}
    \vspace{0.01\linewidth}
    \caption{(a) Model of a coupled cavity OFCG. Again we are simplifying the maths by assuming the crystal is twice as long and light passes it only once per round-trip. (b) Spectral output of the coupled cavity OFCG. Incidental light beam is assumed to have a normalised intensity of 1. Parameters used to obtain this plot were $R_1 = R_2 = 0.99$, $R_3=0.96$, free spectral range of the coupling cavity was set to $21.13$ GHz, while the second cavity was set to match the 5th subharmonic of frequency of modulation $f_m=2.39$ GHz. Modulation depth of the crystal $\delta=\pi/2$, resulting in efficiency of the system of $90.4\%$. (c) Temporal output from the coupled cavity OFCG with a FWHM pulse $\Delta\tau_p = 1.33$ ps as indicated by the red area. Asymmetry is an artifact of the crystal dispersion and finite coupling matrix.}
    \label{2}
\end{figure}

A major limitation of the single cavity OFCG is the trade-off between increasing the modulation depth and the cavity finesse to achieve shorter pulses, and the increased loss from the central cavity mode resulting in an impedance mis-match that inhibits efficient mode-matching into the cavity and hence a low output power \cite{Kovacich:00}. For the parameters presented in previous section the efficiency of the device was predicted to be only $\eta = 0.3\%$, with $<1\%$ typical for experiments \cite{Kovacich:00,Xiao:09}.
\\
This limitation can be overcome using a coupled cavity configuration as demonstrated in \cite{Macfarlane:96} and shown schematically in \fref{2}(a). Here a planar mirror is placed between the input mirror M1 and the EOM crystal surface - to split the OFCG into two separate cavities - an empty coupling cavity (M1 + M2), and a second cavity (M2 + M3) containing the EOM, referred to as the EOM cavity. Here the coupling cavity is designed to be almost lossless, with impedance matched mirrors $R_1=R_2$ to allow for nearly $100\%$ coupling into the EOM cavity.
\\
Using the mathematical model derived in \sref{mathCC} we obtain the steady state of the system by solving the coupled equations
\begin{subequations}
    \begin{gather}
        \Vec{E}_1  = \left[ \mathbbm{1} - \left(r'_1t'_2r_3t_2t_{\mathrm{c}}^2\hat{\phi}_1\times\hat{\phi}_1\times\hat{\phi}\left(\phi_k\right)\times M(2\delta)\times\left\{\mathbbm{1} - r'_2r_3t_{\mathrm{c}}^2\hat{\phi}\left(\phi_k\right)\times M(2\delta)\right\}^{-1} + r'_1r_2\hat{\phi}_1\hat{\phi}_1\right)\right]^{-1}\times t_1\hat{\phi}_1\vec{E}_{\mathrm{in}}, \label{CCEOM:mainbody1}\\
        \vec{E}_{\mathrm{out}}  = t_3t_{\mathrm{c}}\hat{\phi}\left(\frac{\phi_k}{2}\right)\times M(\delta)\left[\mathbbm{1}- r'_2r_3t_{\mathrm{c}}^2\hat{\phi}\left(\phi_k\right)\times M(2\delta)\right]^{-1}\times t_2\Vec{E}_1,\label{CCEOM:mainbody3}
    \end{gather}
\end{subequations}
\noindent where $t_i$ and $r_i$ are the transmitance and reflectivity of the mirror $M_i$, $\hat{\phi_1}$ is a phase change operator representing half round trip phase change between the sidebands in coupling cavity, $\hat{\phi}\left(\phi_k\right)$ is a phase operator representing full round trip phase change between the sidebands in the EOM cavity.   

For the two cavity setup, an additional restriction on the frequency width of the generated comb arises when the $i^\mathrm{th}$ sideband becomes resonant with the $j^\mathrm{th}$ mode of the input coupling cavity. To maximise the comb-width, the FSR of the input cavity is chosen with a large minimum common denominator with the FSR of the EOM cavity, and a large finesse to reduce the cavity linewidth.
The first sideband that is completely resonant with the coupling cavity is called the characteristic mode, which corresponds to the maxiumum sideband order of the comb.

\begin{figure}[t]
    \centering
    \includegraphics[width=.95\textwidth]{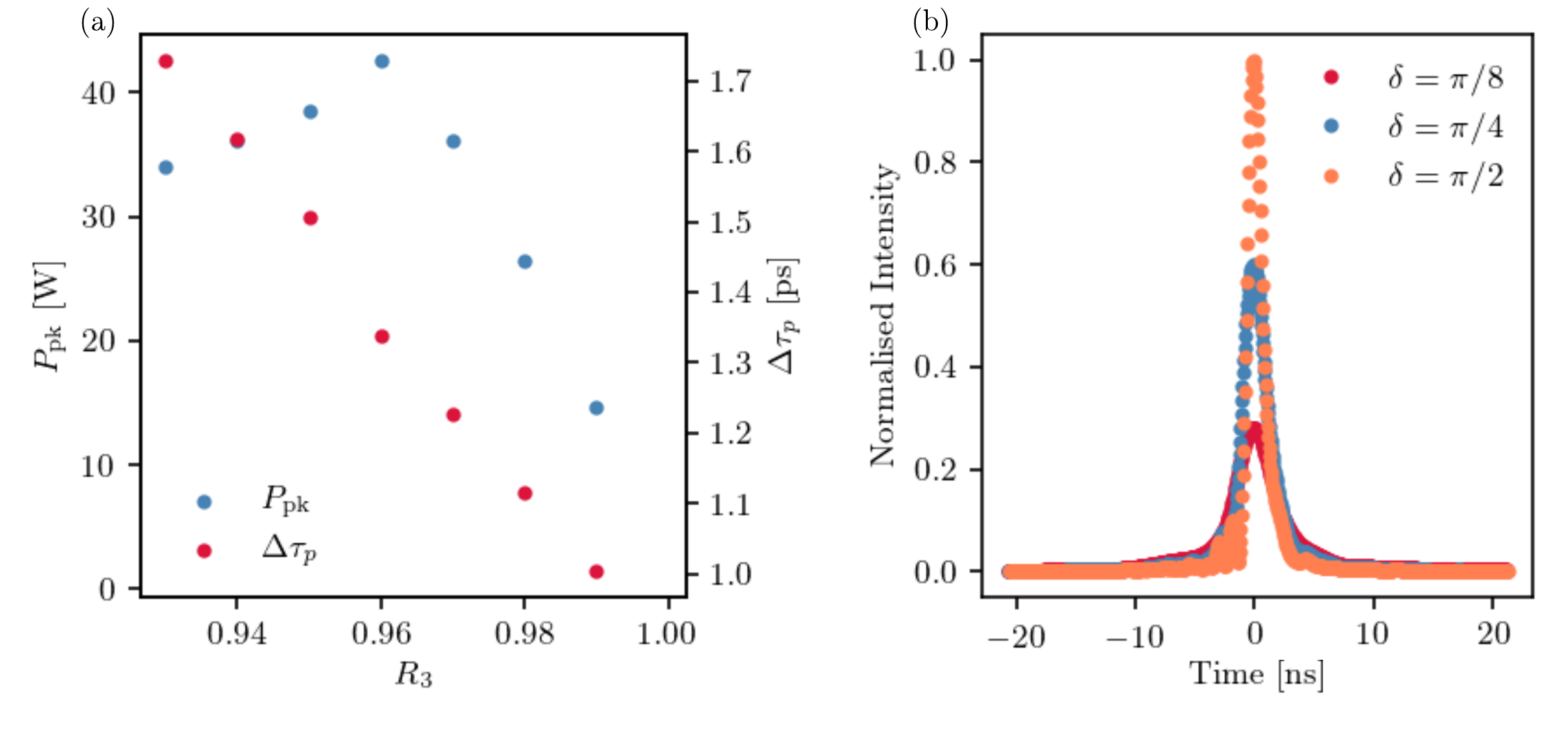}
    \caption{
    (a) Optimisation plot for the Coupled Cavity OFCG. Finesse of the EOM cavity was varied whilst $\delta = \pi/2$, $f_m=$ 2.39 GHz, $\mathrm{FSR_1}=$ 21.13 GHz, $\mathrm{FSR_2}=$ 478 MHz, and $R_1=R_2=0.99$ 
    - the optimal value of the coupling cavity's finesse ($F$ = 300).
    Additionally input power, was scaled so the intra-cavity intensity was below the damage threshold of the EOM crystal. On this plot the relation between peak power, pulse width and finesse can be observed. Increasing the finesse allows for shorter pulses but quickly ramps up the intra-cavity intensity, which places a restriction on $P_{\mathrm{in}}$ and has negative impact on the peak power.
    (b) Variation of modulation depth while the system is operating with $f_m = 2.39$ GHz and for optimised reflectivities of the mirrors ($R_1=R_2=$ 0.99 \& $R_3=$ 0.96). As we decrease the modulation depth below $\pi/2$, the FWHM pulses $\Delta\tau_p$ become longer, respectively 1.3, 2.2 and 3.6 ps, which has negative effect on the peak power.}
    \label{4}
\end{figure}

To enhance photon pair generation using the source, the coupled cavity design was optimised to maximise the peak pulse energy whilst remaining below the damage threshold of the EOM. For the EOM crystal detailed in \sref{SCEOM} to achieve $\delta=\pi/2$ at a modulation frequency $f_m=2.39$ GHz the crystal aperture is limited to $2 \times 2$ mm, with a damage threshold of 20 W/mm$^2$ set by the AR coating. To mitigate this restriction the EOM cavity is extended to set $\mathrm{FSR_2}=f_m/5$, equal to the 5$^{\mathrm{th}}$ subharmonic of the modulation frequency. The source performance is optimised through adjusting mirror reflectivities and radius of curvature to define beam waist in the EOM and the FSR of the input cavity. Possible values of $\mathrm{FSR_1}$ were dictated by the curvatures of the mirrors allowing for desirable beam waist on the crystal facet. Final value of $\mathrm{FSR_1}=21.13$ GHz was chosen due to the characteristic mode appearing farthest away from the central mode in the spectrum. The peak power per pulse was calculated from 
\begin{equation}
    \label{P_p}
    P_{\mathrm{pk}} = \frac{\eta P_{\mathrm{in}}}{\Delta\tau_p2f_m},
\end{equation}
\noindent where $\eta$ is the efficiency of the EOM cycle, $P_{\mathrm{in}}$ is the maximal average pump power of the system (up to 300 mW) - limited by the damage threshold of the EOM crystal. \fref{4}(a) shows results of optimisation of the finesse of the EOM cavity. \fref{4}(b) shows the variation of the modulation depth in the system running with optimised parameters. 

Following the optimisation the coupling cavity is chosen to have $\mathrm{FSR_1}=21.13$ GHz and $R_1=R_2=0.99$ yielding a finesse of 300. The EOM cavity has $\mathrm{FSR_2}=478$ MHz leading to the characteristic mode of 1880 and finesse of 120.
The simulated output field spectrum and temporal profile are shown in \fref{2}(b-c) at the optimised peak power, using system parameters $R_1 = R_2 = 0.99$, $R_3= 0.96$, $\delta = \pi/2$, $f_m = 2.39$ GHz, FSR of the coupling cavity was 21.13 GHz, while the EOM cavity was set to match the $5^{\mathrm{th}}$ subharmonic of frequency of modulation.
For these parameters FWHM of the pulse width $\Delta\tau_p=$ 1.33 ps, $\eta=90.4\%$, the repetition rate equal to 4.78 GHz, peak power $42.4$ W, and beam waist $\omega_0=$ 144 $\mathrm{\mu m}$.



\section{Photon pair generation}
\label{sec:FWM}
\begin{figure}[t]
    \centering
    \includegraphics[width=1\textwidth]{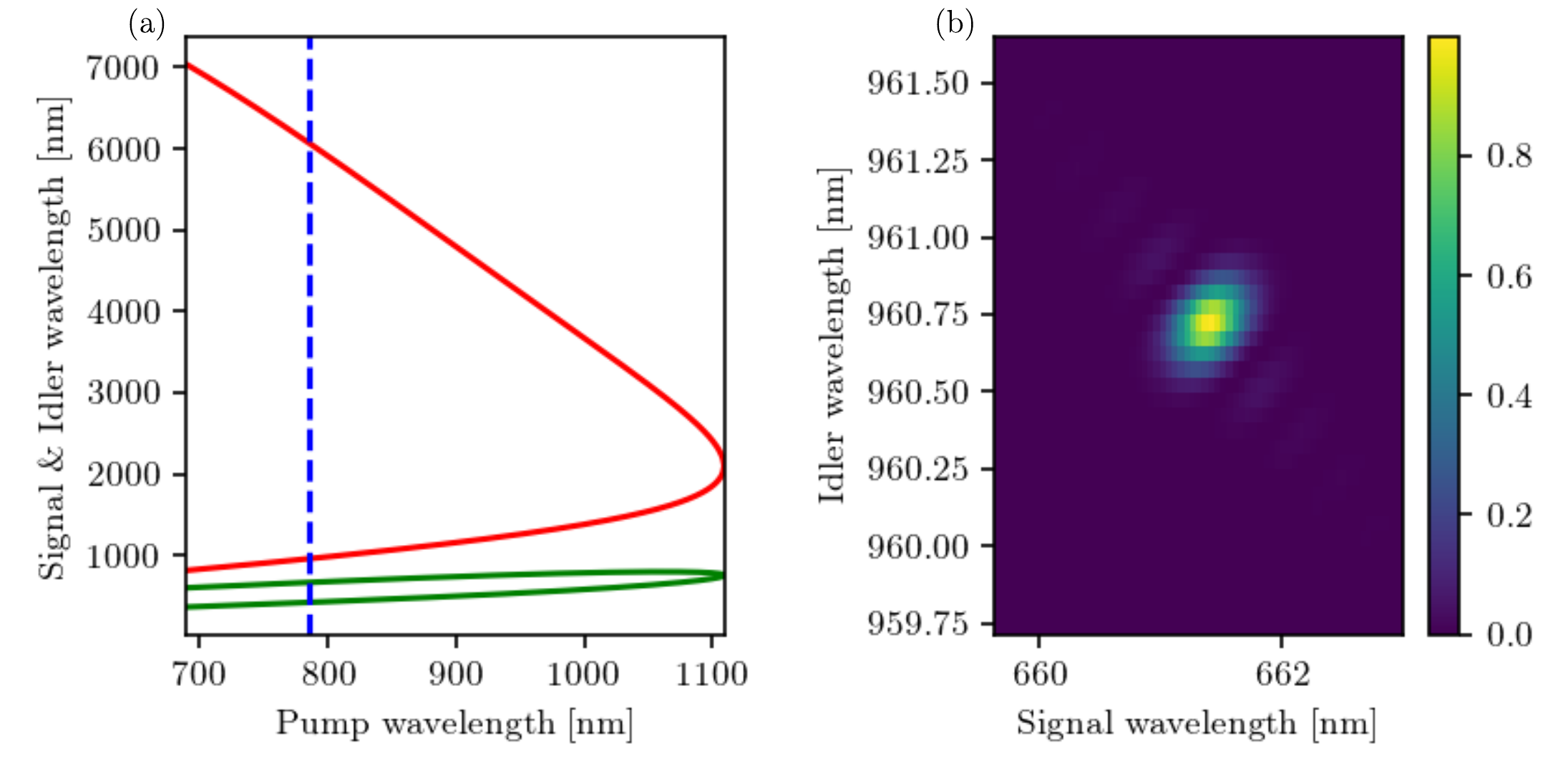}
    \caption{(a) Theoretical birefringent phase-matching contour as a function of the pump central wavelength $\lambda_p$. Signal wavelengths marked in green, idler wavelengths marked in red and blue dotted line marking the central frequency of the pump used. Values for the Fibercore HB800G optical fiber, operating in normal regime $\Delta n=4.3\times 10^{-4}$ \cite{Smith:09}. (b) Resulting joint spectral intensity of the generated signal \& idler pair, with purity of 71\%, fiber is assumed to be $L=10$ cm long. }
    \label{FWM:1}
\end{figure}
For conversion of the output of the coupled cavity OFCG into spectrally pure and factorisable photon pairs for quantum illumination we propose use of FWM in a birefringent fiber, where photon pair frequencies are determined by energy conservation
and by phase matching contour
\begin{equation}
    \label{FWM:eq2}
    \kappa = k_s + k_i - 2k_p + \frac{2}{3}\gamma P_p,
\end{equation}
where, $\kappa$ is the phase-mismatch between the propagation constants of the signal ($s$), idler ($i$) and two pump waves ($p$)  - caused by the chromatic dispersion in the fiber, with $k_j=\frac{n_j\omega_j}{c}$ $\left(j = s, i, p\right)$ for light wave with angular frequency $\omega_j$ propagating in the medium with refractive index $n_j$.

For a HB800G fibre, a 785 nm pump wavelength is converted to a $\sim660$ nm and $\sim960$ nm signal and idler pair - shown in \fref{FWM:1}(a-b), ideally suited for demonstration of quantum LIDAR due to working with conventional GaAs SPAD detectors. 
The spectral purity of the created pair is determined by the joint spectral amplitude (JSA), $f(\omega_s,\omega_i)$\cite{Lugani2020}.
In the ideal case of pure state generation the JSA of the output signal and idler photon pair should be separable - $f(\omega_s,\omega_i) = S(\omega_s)I(\omega_i)$. 
Here functions $S(\omega_s)$ and $I(\omega_i)$ represent the spectral amplitudes of the signal and idler fields.

To quantify the degree of inseparability (the degree of correlation between signal and idler modes) Schmidt decomposition of the JSA can be utilised\cite{doi:10.1080/09500340.2018.1437228}. 
The resulting Schmidt number $K$ represents the number of modes excited in FWM ($K = 1$) for a pure state, which is related to purity via $\mathcal{P} = 1/K$ \cite{doi:10.1080/09500340.2018.1437228}. 
For our system we consider a fiber length 10 cm corresponding to an estimated purity of 71 \% - shown on \fref{FWM:1}(b), which can be increased using shorter fibers at the expense of reduced pair generation rates.

To estimate our expected pair detection rate we scale experimental data from \cite{Smith:09}, 
featuring a commercial pulsed Ti:Sapphire laser generating bandwidth limited pulses with $\Delta\lambda = 3$ nm centred at 704 nm, with a repetition rate $R_r = 80$ MHz.
For an $L_c=$ 10 cm fiber, pair rates of $N_{\mathrm{exp}}=$ 23000 pairs/s are obtained from an average power of 15 mW. Assuming $\mathrm{sech^2}$ pulse shape, this corresponds to pulse width of 0.17 ps with a peak pulse power of $P_{\mathrm{pk,ref}}=$ 1.1 kW.
The pair production rate is linearly proportional to the fiber length L and quadratically proportional to peak power squared, $N\propto LP_{\mathrm{pk}}^2$\cite{PhysRevA.75.023803}. This allows us to estimate the probability of creating a photon pair per pulse with $\alpha = N_{\mathrm{exp}}/R_r,$ which then can be re-scaled using the data from \sref{CCOFCG} to obtain the predicted pair detection rate
\begin{equation}
    \label{pair:prod}
    N = \left(\frac{P_{\mathrm{pk}}}{P_{\mathrm{pk,ref}}}\right)^2 \times \alpha \times 2f_m,
\end{equation}
\noindent where $P_{\mathrm{pk}}$ is peak power per pulse estimated from our model (42.4 W), and $2f_m$ is the repetition rate of our system. We use the above to predict $>2000$ photon pairs detected per second. 
\section{Conclusion}
\label{Sec:Conclusion}
We have presented a highly efficient coupled cavity OFCG optimised for use in generating spectrally pure photon pairs for use in quantum illumination experiments. The source provides a compact, low power and low cost  approach to creating picosecond pulses with high (GHz) repetition rate. We predict using photon pair generation in a commercial birefriengent single mode optical fiber to yield $>2000$ photon pairs/s, with photon pair purity of 71\%.

Unlike commercial pump sources, this design provides a dynamically tunable pulse width through controlling the EOM modulation depth. This method can be adapted to work at other wavelengths and in future extended to telecoms wavelengths. Exploration of the EOM materials offering higher damage thresholds will enable higher peak pulse energy and shorter pulse duration to boost the attainable pair generation rate. The purity of the photon pair generated can be increased by shortening the fibre to $L=6.81$ cm to achieve highest pair purity attainable for this source of 80\% at the cost of a reduction in photon pair production. Alternatively, a longer fiber can be chosen sacrificing purity for the increased pair generation rates.

\acknowledgments 
 
This project is funded by the UK Ministry of Defence. 
\appendix    

\section{Mathematical models}
\subsection{Single Cavity OFCG Mathematical model}
\label{sec:matha}
The output electric field can be calculated by considering the situation in \fref{1}(a).
We are able to represent all of those electric fields as follows

\begin{subequations}
\begin{align}
    \Vec{E}_1 &= t_1\Vec{E}_{\mathrm{in}} + r_1\Vec{E}_2,  \label{eq:eq18a}\\
    \Vec{E}_2 &= t_{\mathrm{c}}^2\hat{\phi}(\phi_k) M(2\delta) \times \Vec{E}_1,\label{eq:eq18b}
    \end{align}
\end{subequations}
\noindent 
where $r_{i}$ and $t_{i}$ stand for reflectivity and transmittance of mirror M$_i$ and $t_{\mathrm{c}}$ is transmittance trough the cavity. Operator $\hat{\phi}(\phi_k)$ is the full round-trip phase change due to crystal dispersion and can be expressed as

\begin{equation}
    \hat{\phi}(\phi_k) = \begin{bmatrix} 
    e^{-i\phi_k} & 0 & \hdots & \hdots & 0 \\
    0 & \ddots & \ddots &   & \vdots \\
    \vdots & \ddots & e^{-i\phi_0} & \ddots & \vdots \\
    \vdots &   & \ddots & \ddots & 0 \\
    0 & \hdots & \hdots & 0 & e^{-i\phi_{-k}}
    \end{bmatrix}, 
\label{eq:eq19}
\end{equation}
\noindent where $k$ is the sideband order and $\phi_k$ is the phase change due to crystal dispersion per sideband, given by \eref{eq:eq17}. $M(2\delta)$ is the transformation matrix of the EOM. 
This matrix can be represented as
\begin{equation}
    M(2\delta)  = \begin{bmatrix} J_0(2\delta) & -J_1(2\delta) & J_2(2\delta) & \hdots & J_{2k}(2\delta)\\
    J_1(2\delta) & J_0(2\delta) & -J_1(2\delta) & \hdots & -J_{2k-1}(2\delta) \\
    J_2(2\delta) & J_1(2\delta) & J_0(2\delta) & \hdots & J_{2k-2}(2\delta) \\
    \vdots & \vdots & \vdots & \ddots & \vdots \\
    J_{2k}(2\delta) & J_{2k-1}(2\delta) & J_{2k-2}(2\delta) & \hdots & J_0(2\delta) 
    \end{bmatrix},
    \label{eq:eq20}
\end{equation}
\noindent where $J_k(\delta)$ is Bessel function identity of order $i$, and an origin $\delta$, and where $\delta$ is the modulation depth.

Using \eref{eq:eq18a} - (\ref{eq:eq18b}) we can solve the field state in $\Vec{E}_1$ depending on incident field $\Vec{E}_{\mathrm{in}}$

\begin{equation}
    \label{eq:eq21}
    \Vec{E}_1 = \left[ \mathbbm{1} - r_1r_2t_{\mathrm{c}}^2\hat{\phi}(\phi_k)\times M(2\delta) \right]^{-1} \times t_1 \Vec{E}_{\mathrm{in}},
\end{equation}
\noindent using this solution, we can calculate output electric field of the OFCG - $\Vec{E}_{\mathrm{out}}$, to be
\begin{equation}
    \Vec{E}_{\mathrm{out}} = t_{\mathrm{c}}\hat{\phi}\left(\frac{\phi_k}{2}\right)\times M(\delta)\times(t_2\Vec{E}_1).
    \label{eq:eq22}
\end{equation}

\subsection{Coupled Cavity OFCG Mathematical Model}
\label{mathCC}
To solve the coupled cavity OFCG we take situation in \fref{2}(a)
\begin{subequations}
\begin{align}
    \Vec{E}_1 & = \hat{\phi}_1\times\left(t_1\Vec{E}_{\mathrm{in}} + r'_1\Vec{E}_4\right),  \label{eq:5.1a}\\
    \Vec{E}_2 & = t_2\Vec{E}_1+r'_2\Vec{E}_3,\label{eq:5.1b}\\
    \Vec{E}_3 & =  t_{\mathrm{c}}^2\hat{\phi}_2\times M(2\delta) \times r_3 \Vec{E}_2, \label{eq:5.1c}\\
    \Vec{E}_4 & =\hat{\phi}_1\times\left(t'_2\Vec{E}_3 + r_2\Vec{E}_1\right),\label{eq:5.1d}
    \end{align}
\end{subequations}
\noindent where $\hat{\phi}_{1}$ is a phase operator, representing half round trip phase change between sidebands, in first cavity, $\hat{\phi}_{2}$ is a phase operator, representing full round trip phase change between sidebands in the EOM cavity, M(2$\delta$) is full round trip transformation matrix, $r(t)_{i}$ represent mirror M$_i$ reflectivity (transmitance) on left side of mirrors respectively, while $r'(t')_{i}$ represent mirror M$_i$ reflectivity (transmitance) on the right side of the mirrors. relation between those can be represented as
\begin{subequations}
\begin{align}
    r' & = r, \label{rprim}\\
    t' & = - t, \label{tprim}
\end{align}
\end{subequations}
\noindent this ensures the transmission phase is preserved.
By substituting \eref{eq:5.1c} into \eref{eq:5.1b} we obtain
\begin{subequations}
\begin{align}
    \Vec{E}_2 & = t_2\Vec{E}_1 + r'_2r_3t_{\mathrm{c}}^2\left(\hat{\phi}_2\times M(2\delta)\times\vec{E}_2\right), \label{eq:5.2a} \\
    \Vec{E}_2 & = \left[\mathbbm{1}- r'_2r_3t_{\mathrm{c}}^2\hat{\phi}_2\times M(2\delta)\right]^{-1}\times t_2\Vec{E}_1, \label{eq:5.2b}
    \end{align}
\end{subequations}
\noindent after that we substitute \eref{eq:5.1c} into \eref{eq:5.1d} to obtain
\begin{equation}
    \Vec{E}_4  = \hat{\phi}_1\left(t'_2r_3t_2t_{\mathrm{c}}^2\hat{\phi}_2\times M(2\delta)\left[\mathbbm{1}- r'_2r_3t_{\mathrm{c}}^2\hat{\phi}_2 M(2\delta)\right]^{-1}\times \vec{E}_1 + r_2\vec{E}_1\right), \label{eq:5.3a} 
\end{equation}
\noindent finally we substitute \eref{eq:5.2b} and \eref{eq:5.3a} into \eqref{eq:5.1a}, to solve the steady state of the system
\begin{equation} 
\Vec{E}_1  = \left[ \mathbbm{1} - \left(r'_1t'_2r_3t_2t_{\mathrm{c}}^2\hat{\phi}_1\times\hat{\phi}_1\times\hat{\phi}_2\times M(2\delta)\times\left\{\mathbbm{1} - r'_2r_3t_{\mathrm{c}}^2\hat{\phi}_2\times M(2\delta)\right\}^{-1} + r'_1r_2\hat{\phi}_1\hat{\phi}_1\right)\right]^{-1}\times t_1\hat{\phi}_1\vec{E}_{\mathrm{in}}, \label{eq:5.4b}
\end{equation}
\noindent now to obtain the output electric field we substitute \eref{eq:5.4b} into \eref{eq:5.2b}, and multiply the result by transmittance of the third mirror, half round trip phase operator in the EOM cavity, and half round trip transformation matrix.
\begin{equation}
    \label{finalcceom}
    \vec{E}_{\mathrm{out}} = t_3t_{\mathrm{c}}\hat{\phi}\left(\frac{\phi_k}{2}\right)\times M(\delta)\vec{E_2}.
\end{equation}
\\

 

\bibliography{report} 
\bibliographystyle{spiebib} 

\end{document}